\documentclass{article}

\usepackage[english]{babel}
\usepackage{subcaption}

\usepackage[letterpaper,top=2cm,bottom=2cm,left=3cm,right=3cm,marginparwidth=1.75cm]{geometry}

\usepackage{amsmath}
\usepackage{graphicx}
\usepackage[colorlinks=true, allcolors=blue]{hyperref}
\usepackage{xcolor}
\def\BibTeX{{\rm B\kern-.05em{\sc i\kern-.025em b}\kern-.08em
    T\kern-.1667em\lower.7ex\hbox{E}\kern-.125emX}}
\newcommand{\dashrule}[1][black]{%
  \color{#1}\rule[\dimexpr.5ex-.2pt]{4pt}{.4pt}\xleaders\hbox{\rule{4pt}{0pt}\rule[\dimexpr.5ex-.2pt]{4pt}{.4pt}}\hfill\kern0pt%
}
\usepackage{booktabs}

\title{A bespoke multigrid approach for magnetohydrodynamics models of magnetized plasmas in PETSc}
\author{Mark F. Adams and Matthew G. Knepley}
\date{}
\begin{document}
\maketitle

\begin{abstract}
Fully realizing the potential of multigrid solvers often requires custom algorithms for a given application model, discretizations and even regimes of interest, despite considerable effort from the applied math community to develop fully algebraic multigrid (AMG) methods for almost 40 years.
Classic geometric multigrid (GMG) has been effectively applied to challenging, non-elliptic problems in engineering and scientifically relevant codes, but application specific algorithms are generally required that do not lend themselves to deployment in numerical libraries.
However, tools in libraries that support discretizations, distributed mesh management and high performance computing (HPC) can be used  to develop  such solvers.

This report develops a magnetohydrodynamics (MHD) code in PETSc (Portable Extensible Toolkit for Scientific computing) with a  fully integrated GMG solver that is designed to demonstrate the potential of our approach to providing fast and robust solvers for production applications.
These applications must, however, be able to provide, in addition to the Jacobian matrix and residual of a pure AMG solver, a hierarchy of meshes and knowledge of the application's equations and discretization.
An example of a 2D, two field reduced resistive MHD model, using existing tools in PETSc that is verified with a ``tilt" instability problem that is well documented in the literature is presented and is an example in the PETSc repository (\path{src/ts/tutorials/ex48.c}).
Preliminary CPU-only performance data demonstrates that the solver can be robust and scalable for the model problem that is pushed into a regime with highly localized current sheets, which generates strong, localized non-linearity, that is a challenge for iterative solvers.

\end{abstract}

PETSc is a widely used numerical library that began in the mid 1990's as a sparse linear algebra solver library \cite{PETSc2022}.
It has since grown to provide a wide array of numerical tools, such as discretizations and distributed mesh management, advanced time integrators, nonlinear solvers, support for downloading and building over 50 third party libraries (solvers, mesh partitioners, adaptive mesh refinement managers, etc.), as well as optimization methods and eigensolvers in the tightly integrated Tao and SLEPc libraries, respectively.
PETSc is a fully distributed memory library, built on the message passing interface (MPI), with support for modern accelerated architectures (``GPUs") for the linear solvers, both built-in solvers and third party solvers, and for GPU sparse matrix assembly.
Support for GPU finite element assembly is currently under development with the libCEED library.
PETSc provides performance portability with two approaches to mitigate risks: Kokkos and Kokkos Kernels for a portable language and low level numerical primitives, respectively, and built-in interfaces to vendor low level numeral primitives.
Note, this approach actually yields three back-ends because Kokkos Kernels can be configured to use vendor ``third party libraries" or built-in methods written in Kokkos.

This paper is intended to demonstrate the potential of a bespoke GMG solver for existing production codes that require solvers for models and discretizations that are not supported well with AMG solvers (eg, hyperbolic problems).
Our approach is to use a modular design in a simple code that can be expanded to support the discretization of a particular application.
The PETSc multigrid framework, and overall support for (matrix) operators, adaptive mesh refinement and discretizations is designed to input users meshes and a user's hierarchy of meshes, and our approach requires that user's provide this grid hierarchy.
Finally, an application would provide the algebraic problem (matrix and right hand side, with a possible matrix-free operator for residual calculations on the fine grid).

This report proceeds by discussing the MHD model and PETSc implementation in \S\ref{sec:model}, introducing the model problem, the 2D ``tilt" instability, showing verification and robustness with thin current sheets, in \S\ref{sec:tilt}, 
presenting small scale CPU performance results in \S\ref{sec:perf}, 
and concluding in \S\ref{sec:conc}.  

\section{The MHD model and PETSc implementation}
\label{sec:model}

This project focuses on MHD applications in fusion energy sciences, such as M3D-C1 and NIMROD.
We implement a two field, 2D, incompressible, resistive MHD code with Poisson brackets and stream functions as per Jardin and Strauss \cite{JARDIN2004133,STRAUSS1998318}.
This approach is designed to be extended to full, two-fluid MHD while reusing all of the terms from the reduced model.
To enforce incompressibility, it is common to introduce stream functions,
 \begin{equation}
  \begin{aligned}
  \mathbf{v}   &= \left(  \frac{ \partial \phi}{\partial y},  -    \frac{ \partial \phi}{\partial x}\right), \\
  \mathbf{B}   &= \left( \frac{ \partial \psi}{\partial y},  -    \frac{ \partial \psi}{\partial x} \right),\\
    \end{aligned}
\end{equation}
where $\mathbf{v}$ is the fluid velocity and $\mathbf{B}$ is the magnetic field.
The two field model uses vorticity $\Omega_z$ and magnetic flux $\psi$, and auxiliary variables potential $\phi$ and (negative) current density $j_z$ according to:
\begin{equation}
  \begin{aligned}
  \partial_t \Omega_z   + \left\{ \Omega_z, \phi \right\}  - \left\{ j_z,  \psi \right\} - \mu \nabla^2_\perp \Omega_z &=  0\\
  \partial_t \psi     + \left\{ \psi, \phi \right\}  - \eta j_z &= 0 \\
\Omega_z  -\nabla^2_\perp\phi    &=  0 \\
  j_z  - \nabla^2_\perp  \psi  &=  0, \\
  \end{aligned}
\end{equation}
with Poisson bracket $\left\{ f, g \right\} = \frac{\partial f}{\partial x}\frac{\partial g}{\partial y} - \frac{\partial f}{\partial y}\frac{\partial g}{\partial x}$. 

The linearization of a bracket $\left\{ f, g \right\}$, given that it is linear in both $f$ and $g$, can be expressed with two Frechet derivatives $D_f\left\{ f, g \right\} + D_g\left\{ f, g \right\} $ where $D_f\left\{ f, g \right\} = \lim_{\epsilon \to 0}\left ( \left\{ f+\epsilon {\hat f}, g \right\} - \left\{ f, g \right\} \right ) / \epsilon =  \left\{ \hat f, g \right\} $, and $D \left\{ f, g \right\} =  \left\{  {\hat f},  g \right\}  + \left\{  f, {\hat g} \right\} $. Each of these two terms are put into the column corresponding to the $\hat \cdot $ term.
The linearized system can be written in matrix form as:
\begin{equation}
L \cdot dU=
\begin{bmatrix}
\frac{1}{ \partial t} -\mu \nabla_\perp^2 + \left \{ \cdot, {\phi} \right \} &  -\left \{ j_z, \cdot \right \} & \left \{ {\Omega_z}, \cdot \right \}  & -\left \{ \cdot,  {\psi} \right \} \\ 
 . & \frac{1}{ \partial t}  + \left \{ \cdot, {\phi} \right \} & \left \{ {\psi}, \cdot \right \} & - \eta M\\ 
 M & . &  -\nabla_\perp^2 & .\\ 
. &  -\nabla_\perp^2 & . & M
 
\end{bmatrix} 
\begin{bmatrix}
{d\Omega_z} \\ 
{d\psi} \\
{d\phi} \\ 
{dj_z}
\end{bmatrix} = \begin{bmatrix}
.\\ 
.\\ 
.\\ 
.\\ 
.
\end{bmatrix}.
\label{eq:2fieldmatrix}
\end{equation}

The PETSc finite element assembly  infrastructure processes the finite element (FE) assembly, calling a user point functions that resembles the strong form.
PETSc supports weak forms with four (hardwired) types of terms with the standard FE inner product $\left (  \cdot,  \cdot \right )$ and test functions $v$:
\begin{itemize}
\item ``g0" mass terms, $M$, of the form $\left (  u,  v \right )$,
\item ``g1" first derivative terms $\left ( \nabla u,  v \right )$,
\item ``g2" is not used herein but it is used for first derivative terms $\left (  u,  \nabla v \right )$,
\item ``g3" for the weak form of a $-\nabla^2 u$ term , $\left ( \nabla u, \nabla v \right )$, after integration by parts, which flips the sign.
\end{itemize}
For scalar problems, the point functions for g0 terms return a scalar, the point functions for g1 and g2 return a vector and the g3 terms return a second order tensor (matrix).
In our case the g3 functions return scaled identity tensors (2x2 matrices) and the mass, g0, functions simply return the scale, 1.0.
The g1 terms generates a vector that scales the test function that is dotted with the basis function gradient in the FE integral internal to PETSc.

For instance, $\left\{ f, \hat g \right\} = \frac{\partial f}{\partial x}\frac{\partial \hat g }{\partial y} - \frac{\partial f}{\partial y}\frac{\partial  \hat g }{\partial x} = \left (\frac{\partial \hat g }{\partial x}, \frac{\partial \hat g }{\partial y}  \right)   \left (-\frac{\partial f }{\partial y}, \frac{\partial f}{\partial x}  \right)^T = \nabla \hat g\cdot \left (-\frac{\partial f }{\partial y}, \frac{\partial f}{\partial x}  \right)^T$.
Our internal naming convention uses `left' and `right' to distinguish between two types of linearized bracket terms and this is a `left' term.
Thus the result of the left g1 functions is $\left (-\frac{\partial f }{\partial y}, \frac{\partial f}{\partial x}  \right)$.
Similarly, the result of the right g1 functions for $\left\{\hat f, g \right\}$ is $\left (\frac{\partial g}{\partial y}, -\frac{\partial g}{\partial x}  \right)$ (this can be seen with the bracket property that switching the order of the arguments simply flips the sign).
For instance, the point function for the $ - \left \{ {j_z}, \cdot \right \} $ term $L(1,2)$ is:
\begin{verbatim}
/* bracket -{ jz , .}, in column of linearized variable .; "n" for negative */
void g1_njz_left(PetscInt dim, PetscInt Nf, PetscInt NfAux, .... , PetscScalar g1[])
{
    PetscInt          i,j;
    const PetscScalar *pJzDer   = &u_x[uOff_x[JZ]]; // gradient of jz_z
    for (i = 0; i < dim; ++i)
      for (j = 0; j < dim; ++j)
        g1[j] += -pJzDer[i]*s_K[i][j]; // This should be unrolled, 1/2 no-ops
}
\end{verbatim}
where  {\tt s\_K} is a DxD skew symmetric matrix that encodes the Poisson bracket (this could be hardwired in future).
The result in {\tt g1} here is $ \left(  \frac{ \partial j_z}{\partial y},  -    \frac{ \partial j_z}{\partial x}\right)^T $.
{\tt s\_K} is defined as:
\begin{equation}
\begin{bmatrix}
0 & 1 \\ 
-1 & 0 \\
\end{bmatrix}.
\end{equation}

\subsection{Numerical methods and reproducibility}
\label{sec:numr}

We discretize (\ref{eq:2fieldmatrix}) with triangular finite elements and use quadratic elements in this work.
Square cells of the cartesian mesh are cut into two triangles.
An initial grid with one cell per processor is generated and PETSc's command line option {\tt -dm\_refine\_hierarchy L}, resulting in a logically and geometrically square subdomain on each processor of size $2^L \times 2^L$.
Standard geometric multigrid preconditioned FGMRES is used as the solver, four iterations of GMRES with diagonal preconditioning is used as the smoother, and a direct solver is used for the coarse grid solver.
This linear solver is used in a standard Newton nonlinear solver iteration.
An adaptive Runge-Kutta (RK2) time integrator is use with a maximum time step of $0.025$ (PETSc command line option {\tt -ts\_type arkimex -ts\_arkimex\_type 1bee -ts\_adapt\_dt\_max 0.025}).
A tolerance of $10^{-5}$ ({\tt -ts\_rtol 1e-5}) resulted in a minimum time step of about $0.0005$, which is in line with the time steps reported by Jardin \cite{JARDIN2004133}.
Run scripts, data output, and reproducibility information are stored in the git repository at \path{gitlab.com/markadams4/mg-m3dc1.git}.

\section{Verification with the tilt instability}
\label{sec:tilt}

The {\it tilt} instability is an unstable MHD equilibrium that is well documented in the literature \cite{JARDIN2004133,STRAUSS1998318,Keppens_2014}.
An initial condition is applied to $\psi$ and is maintained in the boundary condition of the domain $[-2,2]^2$ according to:

\begin{figure}[htbp]
\begin{center}
\includegraphics[width=.5\linewidth]{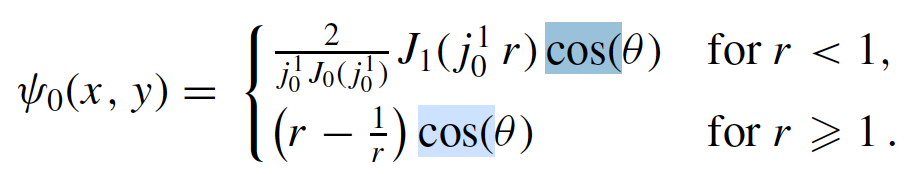}
\label{fig:psi_0}
\end{center}
\end{figure}

This is an MHD equilibrium, but it  is not stable.
One can perturb the initial condition to aid in developing the instability, but we simply allow the instability to develop from round-off errors and perhaps the asymmetry of the mesh (see Figure \ref{fig:tilt2B}).
Strauss provides an image of the initial condition \cite{STRAUSS1998318}, and Jardin provides detailed plots of the evolved fields \cite{JARDIN2004133}.
Jardin conducts extensive verification studies, and our data is in qualitative agreement.
The simulation time of any point (eg, rotation) in the instability depends on the initial conditions and numerical errors.
To compare with Jardin we normalize our time by synchronizing with a point close to Jardin's $t=5$ (Figure 8 (b) and (d) \cite{JARDIN2004133}), which fixes our $\hat t=10.3$ (in the repository \path{gitlab.com/markadams4/mg-m3dc1.git} data file \path{src/data/out_064_data_3_64}) to Jardin's $t=0$.
Note, these plots where generated at intervals of $0.1$ and this synchronization is derived from observation of the plots, and not quantitative data, thus this mapping between our data and Jardin's, $t = \hat t - 10.3$, is approximate.

\begin{figure}[htbp]
\begin{center}
   \begin{subfigure}{0.49\linewidth} \centering
     \includegraphics[scale=0.2]{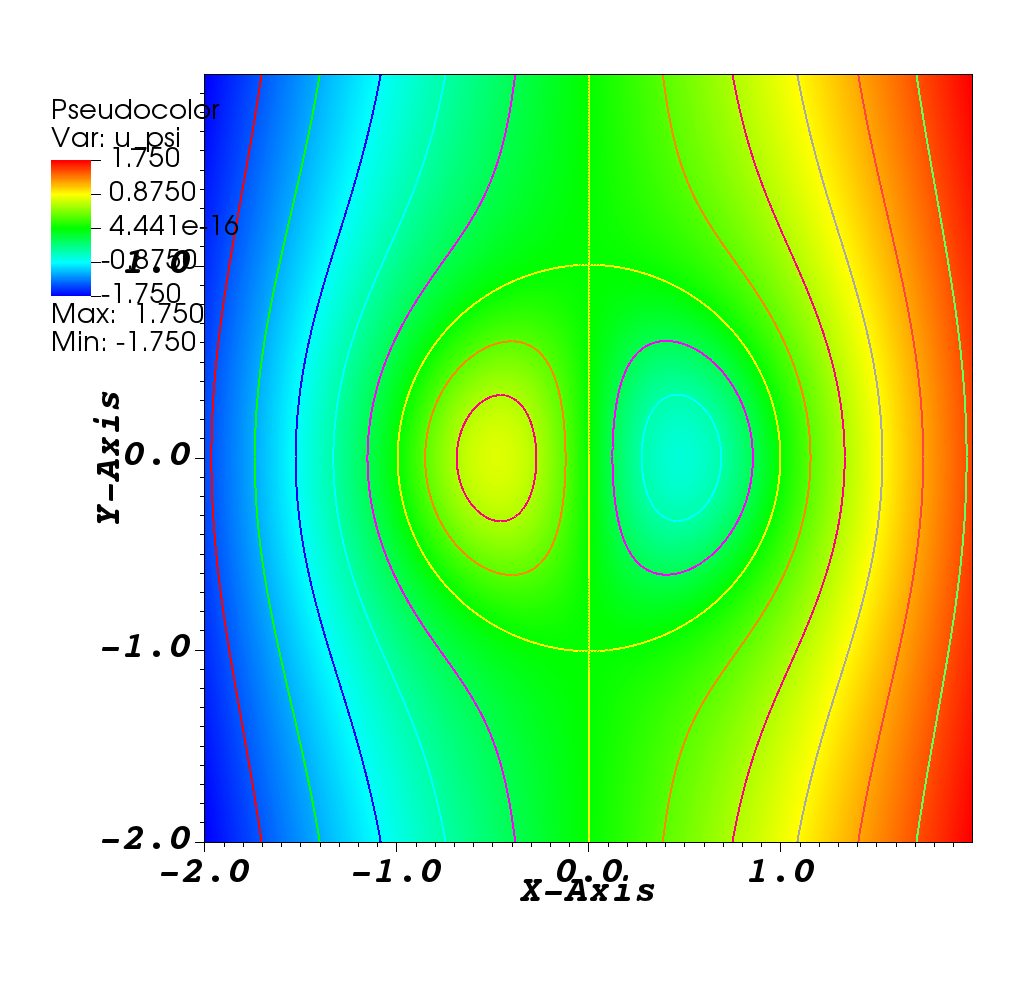}
     \caption{$\psi$}\label{fig:tilt1A}
   \end{subfigure}
   \begin{subfigure}{0.49\linewidth} \centering
     \includegraphics[scale=0.2]{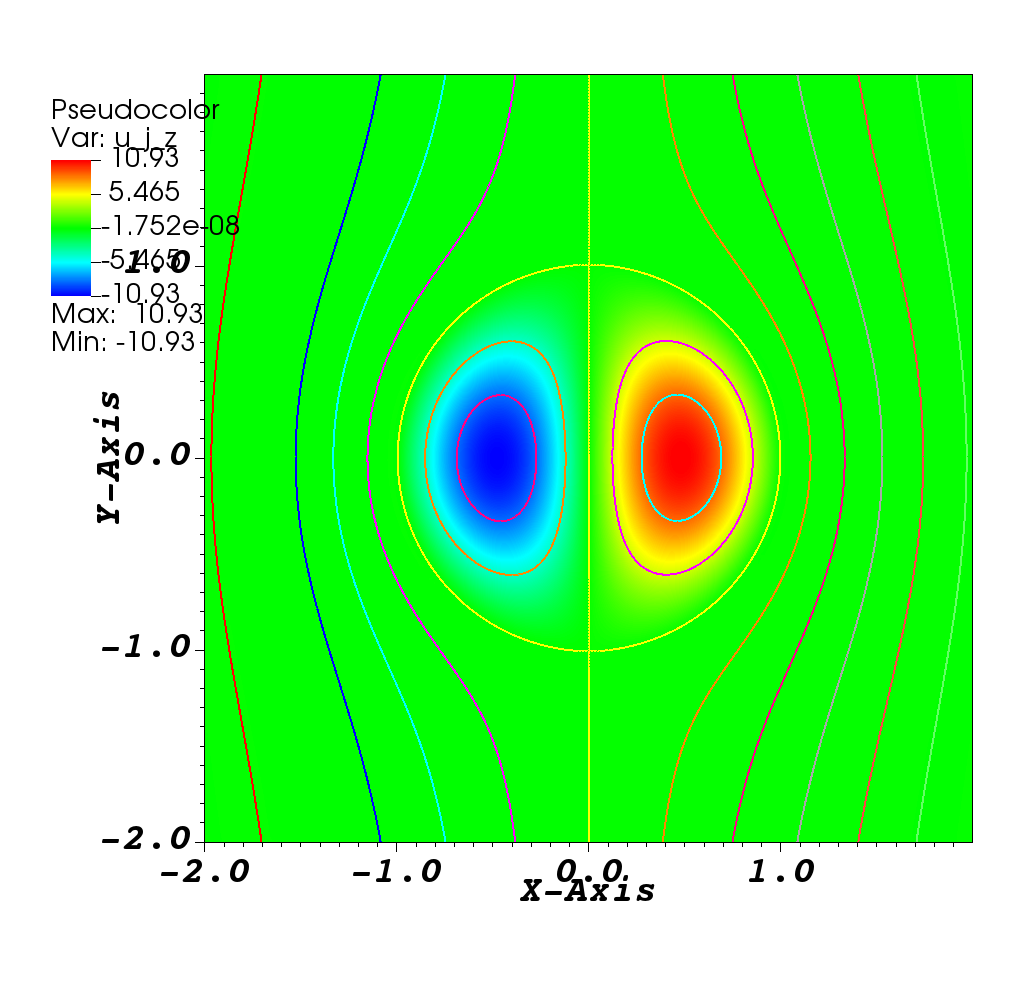}
     \caption{$j_z$}\label{fig:tilt1B}
   \end{subfigure}
\caption{Tilt instability $t=0$ state}
\label{fig:tilta}
\end{center}
\end{figure}

\begin{figure}[htbp]
\begin{center}
   \begin{subfigure}{0.49\linewidth} \centering
     \includegraphics[scale=0.2]{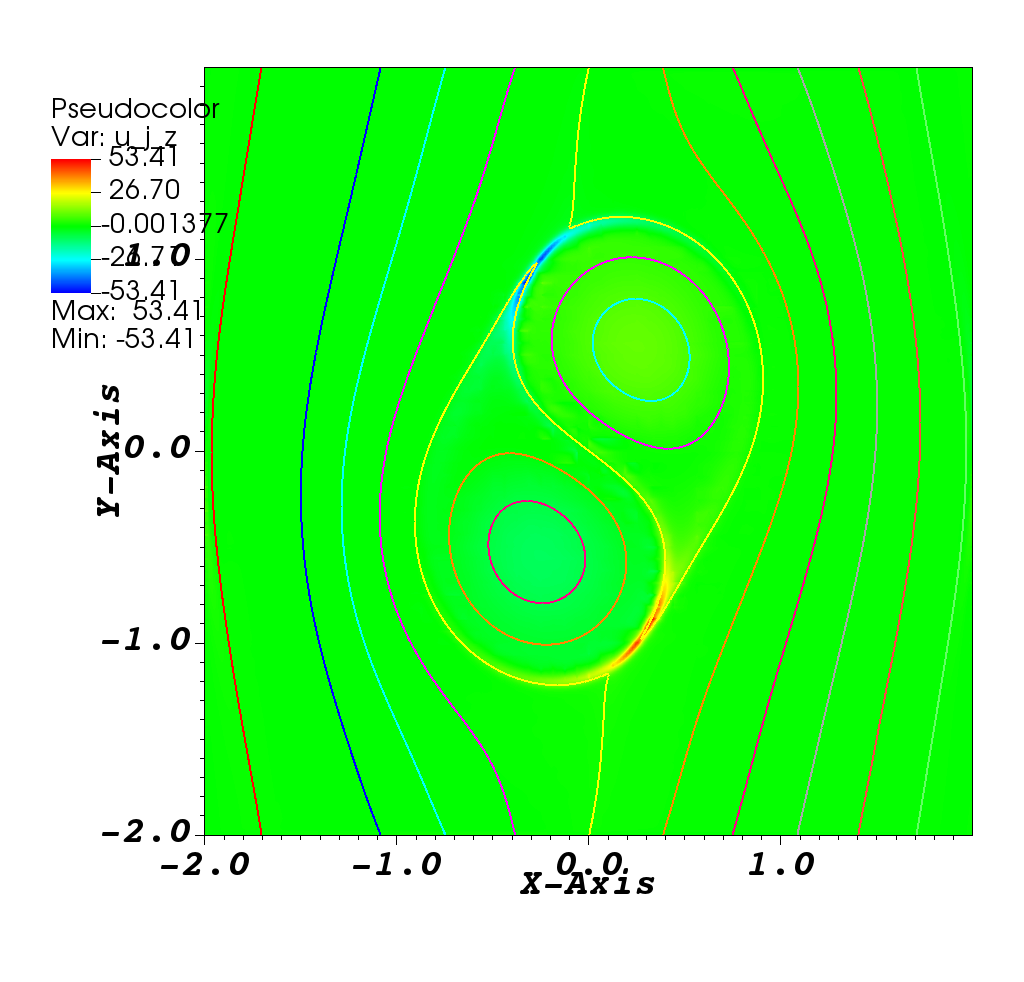}
     \caption{$j_z$}\label{fig:tilt2A}
   \end{subfigure}
   \begin{subfigure}{0.49\linewidth} \centering
     \includegraphics[scale=0.2]{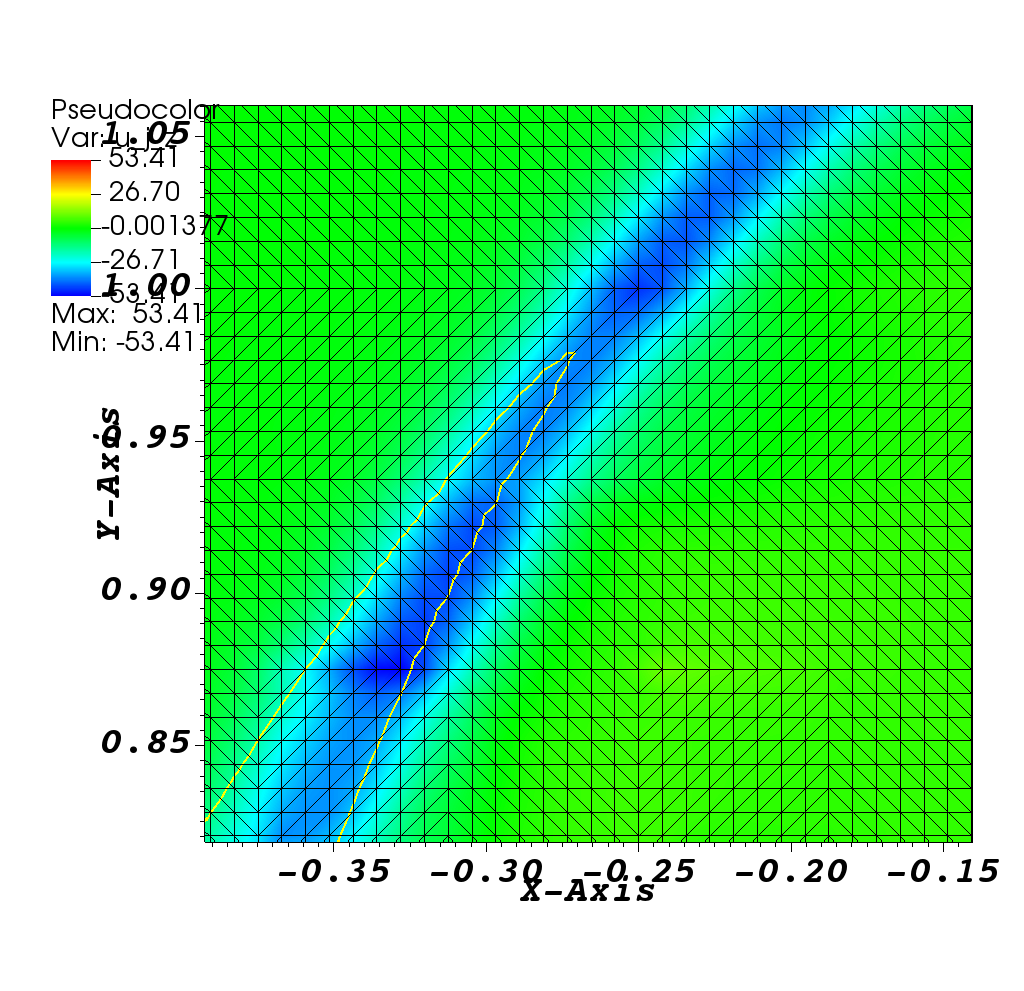}
     \caption{$j_z$ current sheet detail} \label{fig:tilt2B}
   \end{subfigure}
\caption{Tilt instability at time $t=5$ showing some grid effects and visualization artifacts}
\label{fig:tiltb}
\end{center}
\end{figure}

\begin{figure}[htbp]
\begin{center}
   \begin{subfigure}{0.49\linewidth} \centering
     \includegraphics[scale=0.2]{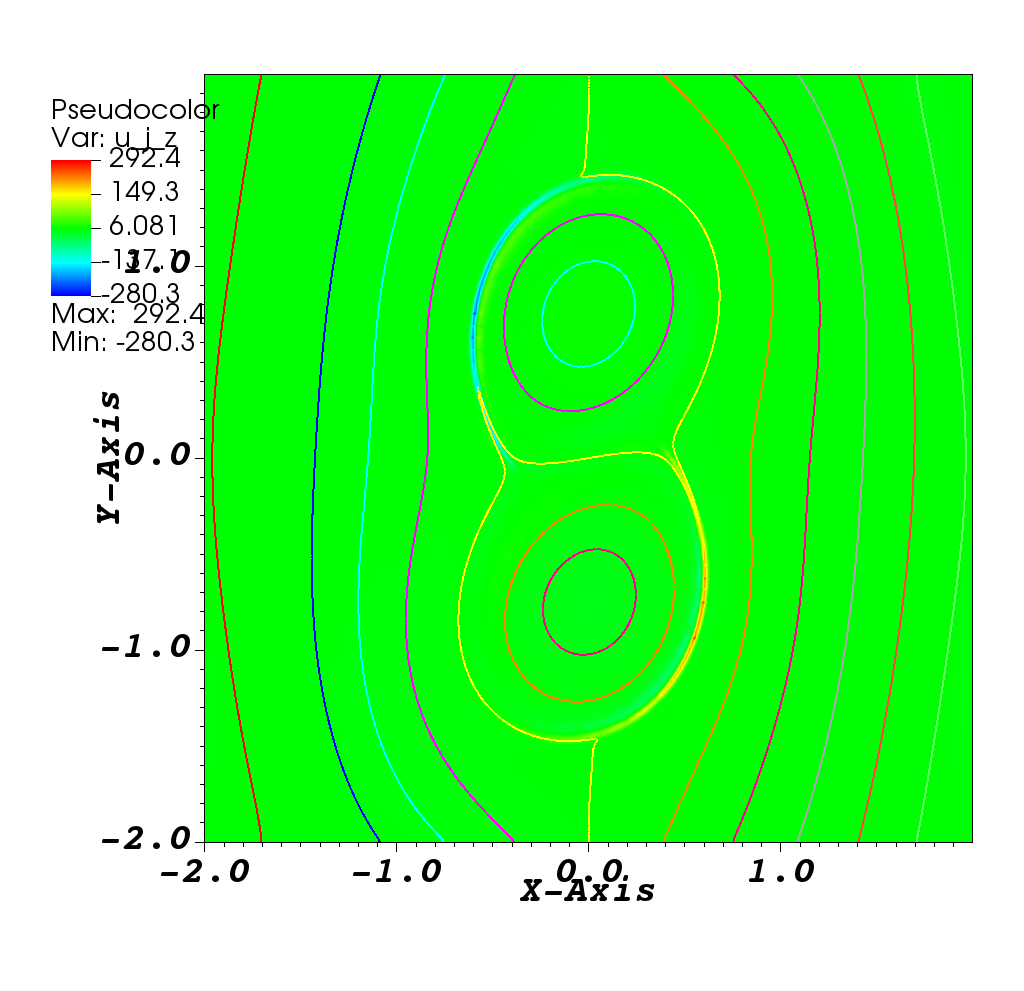}
     \caption{$j_z$} \label{fig:tilt3A}
   \end{subfigure}
   \begin{subfigure}{0.49\linewidth} \centering
     \includegraphics[scale=0.2]{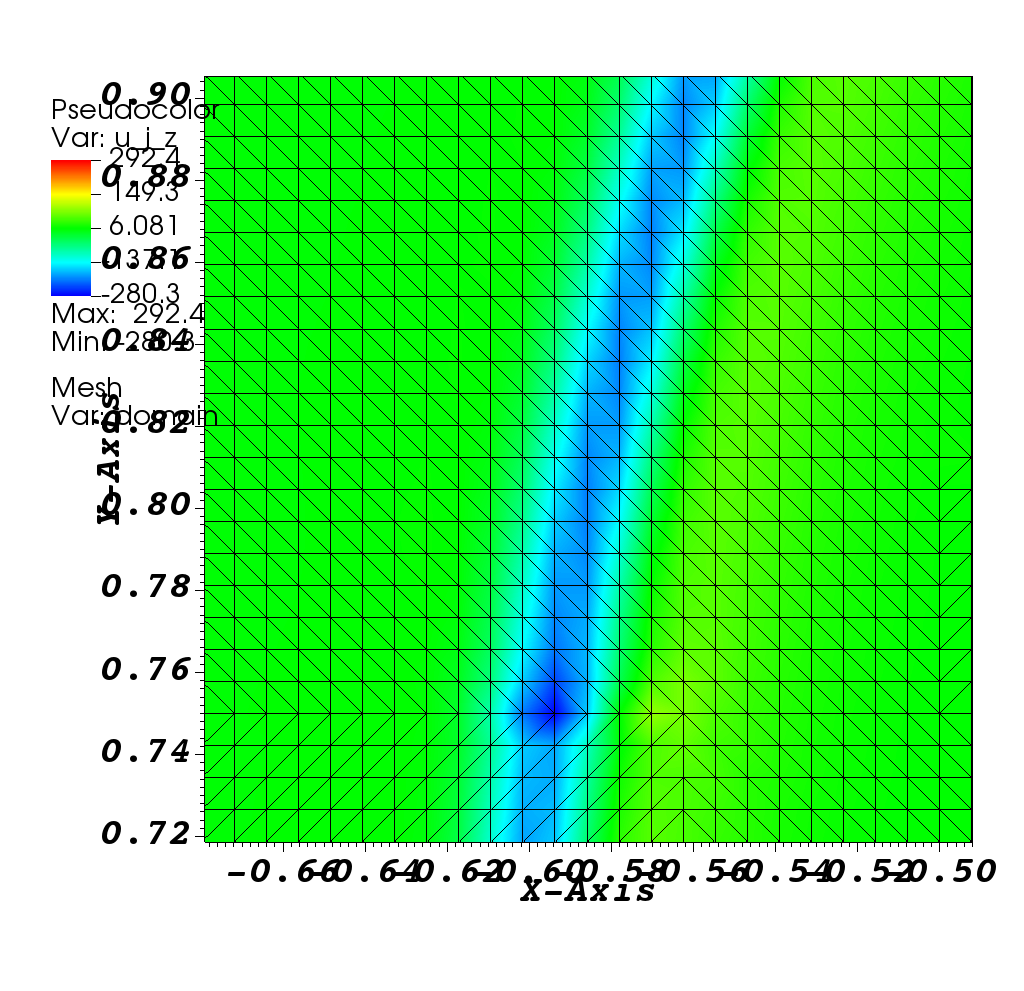}
     \caption{$j_z$ current sheet detail} \label{fig:tilt3B}
   \end{subfigure}
\caption{Tilt instability at time $t=5.8$ with thin current sheets}
\label{fig:tiltc}
\end{center}
\end{figure}

Figures \ref{fig:tilta} -- \ref{fig:tiltc} show plots of the evolution of the test problem with with a $512 \times 512$ cell grid, with each cell simply divided into two triangles.
With $P2$ elements this is produces a $1025 \times 1025$ vertex grid for the solvers with quadratic shape functions.

Figure \ref{fig:tiltb} plots at $t=5$ are in good qualitative agreement with Jardin (Figure 8 (d)) and Figure \ref{fig:tilt2B} shows a detail of the current sheet with some mesh  and visualization effects.
Note, Visit is used for these plots and it uses linear interpolation on a cell and this introduces some of these effects, namely, the dips in the sheet (upper right in Figure \ref{fig:tilt2A}).
Additionally, the direction of the split of cells to generate a triangular mesh is not consistent between processors as is evident in Figures \ref{fig:tilt1B} and \ref{fig:tilt2B}.

Figure \ref{fig:tiltc} plots at the state at $t=5.8$ at a point where the maximum current is observed, although this data is under-resolved there are multiple local maxima are observed.
Jadin compares growth rates qualitatively with previously published results that are consistent with our observed growth rates.

\section{Performance results}
\label{sec:perf}

Performance is evaluated on {\it Crusher} CPU nodes, each with one 64-core AMD EPYC 7A53 “Optimized 3rd Gen EPYC” CPU and two hardware threads per core.
One MPI process is placed on each core, 32 cores are used per node, and the problem is scaled by a factor of two refinement in each dimension for each case.
Two $64 \times 64$ cell subdomains are placed on each core and the node count is increased by a factor of four for each refinement step, maintaining this same size local grid for a weak scaling study with about $32K$ vertices per core or about $1M$ vertices per node, which would be a reasonable if we were using the accelerators.
Note, the finite element device assembly is a work in progress and not available for this study.

Flexible GMRES is used for the solver, preconditioned by a multigrid F-cycle, with 4 Jacobi preconditioned GMRES iterations as the smoother, and PETSc's built-in algebraic multigrid (AMG) solver as the coarse grid solver.
A Newton nonlinear iteration is used with a relative tolerance of $10^{-9}$.
Note, PETSc's AMG solver takes care to reduce the number of active processors on its coarse grids and these solves have a $2\times 2$ cell grid on the coarsest geometric grid, which is the fine grid for the AMG solver, thus, the AMG solver reduces the active processors substantially on the first coarse grid in AMG.
The AMG coarse grid is solve with a serial direct solver on one core.

Table \ref{tab:weak} presents the time in the linear solver for the first time step of the model problem (3 nonlinear solves and a total of 9 linear solves).
\begin{table}[h!]
\centering
\caption{One time step of the three stage adaptive RK2 time integrator with constant subdomain size (weak scaling)} 
\label{tab:weak}
\begin{tabular}{l|ccc}
\toprule
Number of nodes: & 2  & 8 & 32 \\
\midrule
Linear solve times (sec) & 4.0 & 5.9 & 8.0 \\
Total number solver iterations & 24 & 31 & 29 \\
Time per iteration (sec) & 0.17 & 0.19 & 0.28 \\
Number of linear solves & 9 & 9 & 7 \\
\bottomrule
\end{tabular}
\end{table}
This data shows that the implementation scaling is fairly good considering that the multigrid F-cycle applies the coarse grids more often than the traditional V-cycle.
The multigrid F-cycle has the same $O(N)$ work complex, asymptotically, as V-cycle multigrid but the PRAM complexity, or computational depth, increases to $O((log^N)^2)$ from $O(log^N)$ and the constants in the work complexity are a little higher.
These are nonlinear problems and the fluctuation in the iteration count can be caused by new dynamics being resolved, but the source of the fluctuations in iteration counts in this data is not clear.
This data with the full PETSc output is in the repository (\path{gitlab.com/markadams4/mg-m3dc1.git} \path{src/scale/out_0[02|08|32]_scale_6_32})

\section{Conclusion and future work}
\label{sec:conc}

We have presented preliminary results of developing a geometric multigrid solver for the MHD equations is stream function form, with an implementation of the 2-field reduced MHD equations in Jardin and Strauss \cite{JARDIN2004133,STRAUSS1998318}, which demonstrate the feasibility of using PETSc for developing a tightly integrated application and multigrid solver for ideal performance.
We have tested the {\it tilt} instability problem, and run the analysis out to where thin current sheets form that go below the resolution of the mesh used in this study.
The dynamics of the system match those reported in Jardin and the expected physics that we conclude the code is correct.
Preliminary performance results provide only limited data, however the scaling of multigrid methods, while not perfect, is well understood and as good as any solver for a fully couple PDE solve of an elliptic/hyperbolic operator.
These problems are very challenging and we have observed some fragility, like failure with large time steps, which is disappointing but our time steps are in line with Jardin's parameters \cite{JARDIN2004133}.

\subsection*{Future work:}
\begin{itemize}
    \item add block or Venka like smoothers in PETSc with new batch solvers \cite{VANKA1986138,Farrell2019PCPATCHSF};
    \item add this block solver as a new low memory plane solver;
    \item add an anisotropic third dimension and implement semi-coarsening and other multigrid techniques to ameliorate problems from anisotropy to this MHD code;
    \item replace the $P2$ finite elements with a mixed FE method with $C^0$ $\mathbf{B}$ and $\mathbf{v}$ fields as a better approximation to cubic $C^1$ discretization;
    \item investigate the dual discretizations for more stable smoothers;
    \item add device (Kokkos and libCEED) FE device matrix assembly for coarse grid operators;
    \item develop a restriction operator for $C^1$ FE space to our mixed FE space.
\end{itemize}

\section*{Acknowledgments}
This material is based upon work supported by the U.S. Department of Energy, Office of Science, Office of Advanced Scientific Computing Research and Office of Fusion Energy Sciences, Scientific Discovery through Advanced Computing (SciDAC) program.

\bibliographystyle{alpha}
\bibliography{sample}

\newcommand{\etalchar}[1]{$^{#1}$}
\begin{thebibliography}{FKWM19}

\bibitem[BAA{\etalchar{+}}22]{PETSc2022}
Satish {Balay}, Shrirang {Abhyankar}, Mark~F. {Adams}, Steven {Benson}, Jed
  {Brown}, Peter {Brune}, Kris {Buschelman}, Emil {Constantinescu}, Lisandro
  {Dalcin}, Alp {Dener}, Victor {Eijkhout}, Jacob {Faibussowitsch}, William~D.
  {Gropp}, Vaclav {Hapla}, Tobin {Isaac}, Pierre {Jolivet}, Dmitry {Karpeev},
  Dinesh {Kaushik}, Matthew~G. {Knepley}, Fande {Kong}, Scott {Kruger}, Dave~A.
  {May}, Lois~Curfman {McInnes}, Richard~Tran {Mills}, Lawrence {Mitchell},
  Todd {Munson}, Jose~E. {Roman}, Karl {Rupp}, Patrick {Sanan}, Jason {Sarich},
  Barry~F. {Smith}, Stefano {Zampini}, Hong {Zhang}, and Junchao {Zhang}.
\newblock {PETSc: Portable, Extensible Toolkit for Scientific Computation}.
\newblock Astrophysics Source Code Library, record ascl:2210.016, October 2022.

\bibitem[FKWM19]{Farrell2019PCPATCHSF}
P.~E. Farrell, M.~G. Knepley, F.~W., and L.~Mitchell.
\newblock {PCPATCH}: software for the topological construction of multigrid
  relaxation methods.
\newblock {\em ACM Trans. Math. Softw.}, 47:25:1--25:22, 2019.

\bibitem[Jar04]{JARDIN2004133}
S.C. Jardin.
\newblock A triangular finite element with first-derivative continuity applied
  to fusion mhd applications.
\newblock {\em Journal of Computational Physics}, 200(1):133--152, 2004.

\bibitem[KPX14]{Keppens_2014}
R.~Keppens, O.~Porth, and C.~Xia.
\newblock Interacting tilt and kink instabilities in repelling current
  channels.
\newblock {\em The Astrophysical Journal}, 795(1):77, oct 2014.

\bibitem[SL98]{STRAUSS1998318}
H.R. Strauss and D.W. Longcope.
\newblock An adaptive finite element method for magnetohydrodynamics.
\newblock {\em Journal of Computational Physics}, 147(2):318--336, 1998.

\bibitem[Van86]{VANKA1986138}
S.P Vanka.
\newblock Block-implicit multigrid solution of navier-stokes equations in
  primitive variables.
\newblock {\em Journal of Computational Physics}, 65(1):138--158, 1986.

\end{thebibliography}

\end{document}